\documentclass[11pt,preprint,flushrt]{aastex}
\slugcomment{\it Accepted to the Astronomical Journal}
\lefthead{Stern et al.}
\righthead{Mid-Infrared Colors of EROs}
 
\def\eg{{e.g.~}}

\def\hr{HR\,10}
\def\lbds{LBDS\,53W091}
\def\lbdsb{LBDS\,53W069}
\def\deg{\ifmmode {^{\circ}}\else {$^\circ$}\fi}
\def\h70{$h_{70}^{-1}$}
\def\kmsMpc{\ifmmode {\rm\,km\,s^{-1}\,Mpc^{-1}}\else
    ${\rm\,km\,s^{-1}\,Mpc^{-1}}$\fi}
\def\spose#1{\hbox to 0pt{#1\hss}}
\def\simlt{\mathrel{\spose{\lower 3pt\hbox{$\mathchar"218$}}
     \raise 2.0pt\hbox{$\mathchar"13C$}}}
\def\simgt{\mathrel{\spose{\lower 3pt\hbox{$\mathchar"218$}}
     \raise 2.0pt\hbox{$\mathchar"13E$}}}
 
\def\plotfiddle#1#2#3#4#5#6#7{\centering \leavevmode
\vbox to#2{\rule{0pt}{#2}}
\includegraphics{#1}}
 
\begin{document}
 
\title{{\it Spitzer} Observations of the Prototypical Extremely Red Objects \\
HR\,10 and LBDS\,53W091: Separating Dusty Starbursts from Old Elliptical Galaxies}

\author{Daniel~Stern\altaffilmark{1},
Ranga-Ram~Chary\altaffilmark{2}, \\
Peter~R.~M.~Eisenhardt\altaffilmark{1} \&
Leonidas~A.~Moustakas\altaffilmark{1}}

\altaffiltext{1}{Jet Propulsion Laboratory, California Institute of
Technology, Pasadena, CA 91109; {\tt stern@zwolfkinder.jpl.nasa.gov}}

\altaffiltext{2}{{\it Spitzer} Science Center, California Institute of
Technology, Pasadena, CA 91125}

\begin{abstract}

We present {\it Spitzer Space Telescope} observations of the well-studied
extremely red objects (EROs) \hr\ and \lbds\ from 3.6~$\mu$m to
160~$\mu$m.  These galaxies are the prototypes of the two primary classes
of EROs:  dusty starbursts and old, evolved galaxies, respectively.  Both
galaxies, as well as \lbdsb, another example of an old, quiescent galaxy,
are well-detected out to 8 $\mu$m.  However, only the dusty starburst
\hr\ is detected in the far-infrared.  All three EROs have stellar masses
of a few $\times\, 10^{11} M_\odot$.  Using evolutionary model fits to
their multiband photometry, we predict the infrared colors of similar
EROs at $1 < z < 2$.  We find that blueward of observed 10~$\mu$m,
the two ERO classes are virtually indistinguishable photometrically.
Deep spectroscopy and 24~$\mu$m data allow the classes to be separated.

\end{abstract}


\section{Introduction\label{sec:intro}}

It has now been nearly twenty years since the first near-infrared
surveys identified an intriguing population of optically faint galaxies
with surprisingly bright near-infrared magnitudes \markcite{elston:88,
elston:89}(Elston, Rieke, \& Rieke 1988, 1989). Initially proposed to be extremely distant ($z \sim 10$)
sources whose optical emission was absorbed by intergalactic hydrogen
\markcite{Mobasher:05}(\eg Mobasher {et~al.} 2005), the ``extremely red object'' (ERO)
population\footnote{EROs are typically selected to have
  extreme optical-minus-infrared colors, such as $(R-K)_{\rm Vega}>6.0$.} is now
instead recognized to be primarily comprised of two interesting galaxy
populations: (1) old, evolved galaxies whose red colors are caused by
a dearth of young, luminous, massive stars, and (2) dusty starburst
galaxies whose red colors are caused by dust absorption of the bluer
light in actively star-forming regions.  The prototype of each class
is, respectively, the ``old, dead, and red'' radio galaxy \lbds\
at $z=1.552$ \markcite{dunlop:96, spinrad:97}(Dunlop {et~al.} 1996; Spinrad {et~al.} 1997) and the dusty starburst
galaxy \hr\ at $z=1.44$ \markcite{hu:94, graham:96, dey:99}(Hu \& Ridgway 1994; Graham \& Dey 1996; Dey {et~al.} 1999).  The weak
radio galaxy \lbdsb\ at $z=1.432$ \markcite{nolan:03, dey:06}(Nolan {et~al.} 2003; Dey {et~al.} 2006) is a second
well-studied example of the old, evolved ERO class.  EROs have a surface
density of approximately 1000\,deg$^{-2}$ for $R-K>6$ and $K<20$ (Vega),
and comprise $\sim 10\%$ of all $K$-selected galaxies to $K\simlt20$
\markcite{thompson:99, Moustakas:04}(Thompson {et~al.} 1999; Moustakas {et~al.} 2004).  For a recent comprehensive review
of EROs see \markcite{mccarthy:04}McCarthy (2004).

With redshifts $1 \simlt z \simlt 2$ \markcite{cimatti:03}(\eg Cimatti
{et~al.} 2003), clustering properties similar to elliptical galaxies
\markcite{daddi:03, moustakas:02}(Daddi {et~al.} 2003; Moustakas
\& Somerville 2002), and substantial stellar masses that contribute
significant fractions of the global stellar mass density at the redshifts
they are found \markcite{kong:06}(\eg Kong {et~al.} 2006), EROs clearly
are an important population in the context of understanding galaxy
formation and evolution \markcite{somerville:04, nagamine:05}(\eg
Somerville {et~al.} 2004; Nagamine {et~al.} 2005).  Volume-limited
galaxy surveys of the local universe find a clear bimodality in galaxy
properties, with the ``red sequence'' generally populated by the more
massive and established galaxies \markcite{kauffmann:03, blanton:05}(\eg
Kauffmann {et~al.} 2003; Blanton {et~al.} 2005).  There is increasing
evidence that this sequence is securely in place not only out to $z\sim1$
\markcite{bell:04, faber:06}(Bell {et~al.} 2004; Faber {et~al.} 2006),
but possibly to much higher redshifts \markcite{RoccaVolmerange:04,
Labbe:05, papovich:06}({Rocca-Volmerange} {et~al.} 2004; Labb\'e {et~al.}
2005; Papovich {et~al.} 2006).  The processes by which these massive
galaxies form must happen early.  This is consistent with measurements
of the evolution of the global stellar mass density $\Omega_{*}(z)$
in field surveys.  While between half and three-quarters of the
present-day stellar mass is in place by $z\sim1$ \markcite{dickinson:03,
rudnick:03, fontana:04, drory:04}(Dickinson {et~al.} 2003; Rudnick
{et~al.} 2003; Fontana {et~al.} 2004; Drory {et~al.} 2004), only
$\sim 10\%$ of the present-day stellar mass is in place at $z \simgt 3$
\markcite{drory:05}(Drory {et~al.} 2005).  The evolution of $\Omega_{*}$
is dramatic: either the high-redshift accountings are incomplete
(\eg because of cosmic variance effects, unaccounted for obscured
populations, or significant population synthesis deficiencies), or
probes in that redshift range will see the most dramatic epoch of galaxy
assembly in progress.  Indeed, studies of infrared selected samples
of distant galaxies in the Great Observatories Origins Deep Survey
\markcite{Giavalisco:04}(GOODS; Giavalisco {et~al.} 2004)
find significant evidence of massive galaxy assembly at $z \sim 1 -
3$ \markcite{Caputi:06, papovich:06}({Caputi} {et~al.} 2006; Papovich
{et~al.} 2006).  Understanding ERO demographics and energetics will
provide key insight into all of these fundamental questions.

EROs are represented by two different galaxy populations.  With enough
dedicated time and effort, young, dusty starbursts and old, evolved
galaxies may be distinguished through spectroscopic features in the
restframe UV and optical \markcite{Cimatti:02}(\eg Cimatti {et~al.} 2002a).  A photometric
technique proposed by \markcite{Pozzetti:00}Pozzetti \& Mannucci (2000) to distinguish between ERO
types using combinations of optical and near-infrared colors was designed
to work for objects around $z\approx1.5$ \markcite{Mannucci:02}(\eg Mannucci {et~al.} 2002).
\markcite{Dorman:03}Dorman, {O'Connell}, \& Rood (2003) suggest an alternate discriminating technique using
mid-ultraviolet colors.  A more detailed look at the actual $\lambda_{\rm
rest}\sim0.1-1.1\, \mu$m spectral energy distributions (SEDs) of EROs
by \markcite{Moustakas:04}Moustakas {et~al.} (2004) demonstrates that a larger baseline, extending
longward of the observed $K$-band, is needed to differentiate all but
the most dramatic EROs.

In this paper we present 3.6\,$\mu$m to 160\,$\mu$m observations of
the archetypical EROs obtained with the {\it Spitzer Space Telescope}
\markcite{werner:04}(Werner {et~al.} 2004).  Section~2 briefly summarizes previous studies of
these three EROs.  Section~3 describes the {\it Spitzer} observations
and data reduction, followed by an analysis of the broadband SEDs
of these galaxies in \S~4.  Surprisingly, the two ERO populations
remain nearly indistinguishable out to observed 8\,$\mu$m (rest-frame
3.2\,$\mu$m) and it is only at longer wavelengths that the 
populations separate (\S~5).  We adopt the concordance cosmology and use
Vega-system magnitudes unless stated otherwise.

% A companion paper (XXXX et al., in preparation) discusses the
% implications of these results with respect to EROs detected in the
% Groth Strip.

% We adopt a flat cosmology with $\Omega_M = 1 - \Omega_{\Lambda} = 0.3$
% and $H_0 = 70~ h_{70}~ \kmsMpc$, $\Omega_0 = 1$ throughout this paper.

% [{\it possible brief description of qso-2's and EROs?  qso-2's have red
% colors (e.g., lyxn qso-2), but are a small fraction of the ERO
% population.  discuss here?  conclusions?  at all? --- i'm inclined to
% skip.}]

\section{Prototypical EROs\label{sec:archetype}}

\subsection{\hr}

\hr\ (or ERO~J164502$+$4626.4) is the archetype of the dusty starburst
subclass of EROs.  With $K' = 18.4$ and $I - K = 6.0$, \hr\ was
first identified by \markcite{hu:94}Hu \& Ridgway (1994) in deep,
multiband imaging of the damped Ly$\alpha$ quasar PC~1643$+$4631A
\markcite{schneider:91a}($z = 3.79$; Schneider, Schmidt, \& Gunn 1991).
\markcite{hu:94}Hu \& Ridgway (1994) initially suggested that \hr\
and another ERO in the field (HR\,14 or ERO~J164457$+$4626.0) could be
distant, evolved ellipticals lying at $z \sim 2 - 3$, but subsequent
observations from the Keck telescopes by \markcite{graham:96}Graham
\& Dey (1996) and \markcite{dey:99}Dey {et~al.} (1999) showed that
\hr\ is a moderately distant ($z = 1.44$) galaxy with an asymmetric
morphology and [\ion{O}{2}] and H$\alpha$ in emission.  This suggested
that \hr\ was not related to an evolved elliptical, but was rather
a distant counterpart of the local ultraluminous infrared galaxies
(ULIRGs) discovered by the {\em Infrared Astronomical Satellite}
({\em IRAS}).  Strong detections of \hr\ at submillimeter wavelengths
\markcite{cimatti:98b, dey:99, greve:03}(Cimatti {et~al.} 1998; Dey
{et~al.} 1999; Greve, Ivison, \& Papadopoulos 2003) showed the presence of
significant dust \markcite{greve:03}($M_{\rm dust} \sim 9 \times 10^{8}~
M_\odot$; Greve {et~al.} 2003), unambiguously demonstrating that \hr\
is indeed a gas-rich, dust-enshrouded galaxy.

A long-standing open issue has been the role of a possible active nucleus
in the (long-wavelength) energetics of \hr.  Assuming the submillimeter
flux is mainly due to optically-thin thermal emission from dust heated
by a young, star-forming population, the $8 - 1000\, \mu$m luminosity,
$L_{\rm FIR} \sim 9 \times 10^{12}~ L_\odot$, translates to a star
formation rate of about $900~ M_\odot~ {\rm yr}^{-1}$ \markcite{greve:03}(Greve {et~al.} 2003).
This starburst-dominant hypothesis has been favored by the lack of strong
evidence of an AGN in the spectral data \markcite{dey:99}(Dey {et~al.} 1999).  Furthermore, the
mid-infrared SED of \hr\ as measured by ISOCAM on the {\em Infrared Space
Observatory} ({\em ISO}) is similar to that of Arp~220, a well-studied,
local ULIRG whose dominant source of energy is believed to be star
formation \markcite{elbaz:02b}(Elbaz {et~al.} 2002).

\begin{figure}
\figurenum{1a}
\begin{center}
\plotfiddle{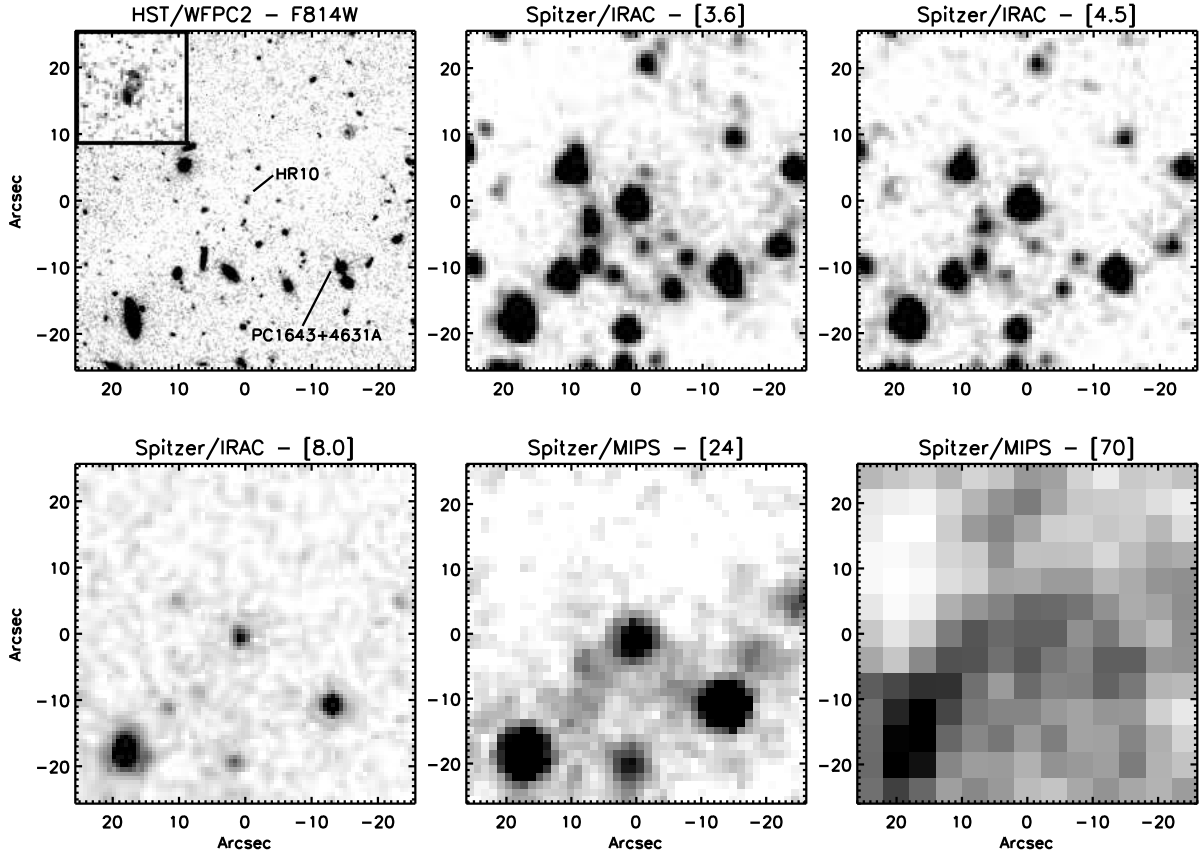}{4.2in}{0}{75}{75}{-240}{-20}
\end{center}

\caption{{\it HST} and {\it Spitzer} images of the prototypical ERO
\hr, located at $\alpha = 16^{\rm h}45^{\rm m}02.36^{\rm s}, \delta =
+46\deg26\arcmin25\farcs5$ (J2000).  Cameras and filters are indicated.
The fields of view shown are 50\arcsec\ on a side and relevant sources
are labeled.  North is up, and east is to the left.  The inset is
4\arcsec\ on side.  Note the galaxy approximately 30\arcsec\ SE of \hr\
which dominates at $70~ \mu$m.  This source also dominates the $160~
\mu$m image; we suggest that it may be contaminating the published $850~
\mu$m photometry as well.}

\label{fig.image_hr10}
\end{figure}

The $850 \mu$m photometry of \hr\ remains somewhat uncertain, with several
Submillimetre Common-User Bolometer Array \markcite{Holland:99}(SCUBA; Holland {et~al.} 1999)
observations resulting in disparate flux density measurements (see
Table~\ref{table.phot_hr10}).  The observations by \markcite{cimatti:98b}Cimatti {et~al.} (1998)
and \markcite{dey:99}Dey {et~al.} (1999) were both obtained using SCUBA in ``photometry
mode'', providing a single pixel measurement.  The latter data
set is twice as large and was obtained in exceptional observing
conditions and is therefore presumed the more reliable measurement.
The observations by \markcite{greve:03}Greve {et~al.} (2003) were obtained in ``jiggle map
mode'', providing an image of the field and photometry closer to that
measured by \markcite{cimatti:98b}Cimatti {et~al.} (1998).  \markcite{greve:03}Greve {et~al.} (2003) identify a $6 \pm
2$~mJy SCUBA source approximately 50\arcsec\ SSW of \hr, and suggest
that the extended emission from this second source may have affected the
sky background estimates in the photometry mode observations, thereby
corrupting those measurements.  An additional source, approximately
30\arcsec\ SE of \hr, may also be confusing the SCUBA measurements.
This source is faintly visible in the jiggle map of \markcite{greve:03}Greve {et~al.} (2003),
and is well-detected in our {\em Spitzer} observations described in \S~3,
becoming significantly brighter than \hr\ at wavelengths longward of $50~
\mu$m (see Fig.~\ref{fig.image_hr10}).

\subsection{\lbds}

\lbds\ is the archetype of the evolved elliptical subclass of EROs.  With
$K = 18.7$ and $R - K = 5.8$, \lbds\ is among the reddest counterparts
to weak radio sources (1~mJy $< S_{\rm 1.4 GHz} < 50$~mJy) in the
Leiden-Berkeley Deep Survey \markcite{Windhorst:84a, Windhorst:84b}(LBDS; Windhorst, {van~Heerde}, \&  Katgert 1984b; Windhorst, Kron, \&  Koo 1984a).
Probing sources with these faint radio fluxes is expected to identify
sources whose scattered AGN contribution to the total continuum emission
is small \markcite{dunlop:93, eales:93}(\eg Dunlop \& Peacock 1993; Eales \& Rawlings 1993).  Spectroscopic observations
with the Keck telescope provided an absorption-line redshift, $z = 1.552$,
and a rest-frame ultraviolet spectrum very similar to an F6~V star and the
local elliptical galaxy M32 \markcite{dunlop:96, spinrad:97}(Dunlop {et~al.} 1996; Spinrad {et~al.} 1997).  The optical
spectrum shows no evidence of either an AGN or recent star formation.

The implied stellar age of $\geq 3.5$~Gyr was cosmologically-interesting
and controversial at the time of initial publication
\markcite{spinrad:97, krauss:97, yi:00}(\eg Spinrad {et~al.} 1997; Krauss 1997; Yi {et~al.} 2000), which was prior to the
current ``concordance cosmology.''  For example, an $\Omega_m = 1$ and
$H_0 = 70 \kmsMpc$ Einstein-de~Sitter universe is only 2.3~Gyr old at
$z = 1.552$.  Even for the ``concordance cosmology'' ($\Omega_{\rm M}
= 1 - \Omega_\Lambda = 0.3$ and $H_0 = 70~ \kmsMpc$), the universe
is only 4.1~Gyr old at $z = 1.552$, requiring an early formation of
the stellar populations of \lbds.  Galaxy~3a, approximately 3\arcsec\
SE of \lbds\ and a few tenths of a magnitude fainter, has an optical
spectrum and optical to near-IR colors similar to \lbds.  Galaxy~3a is
thought to be a faint companion, of similar age and redshift to \lbds\
\markcite{spinrad:97, Bunker:06}(Spinrad {et~al.} 1997; Bunker {et~al.} 2006).

Images of \lbds\ with the Near-Infrared Camera and Multiobject
Spectrograph (NICMOS) on the {\em Hubble Space Telescope} ({\em HST})
reveal \lbds\ to have a regular $r^{1/4}$ \markcite{deVaucouleurs:48}{de~Vaucouleurs} (1948)
profile with an effective radius of $0\farcs32 \pm 0\farcs08$ \markcite{waddington:02, Bunker:06}($2.7
\pm 0.7$ kpc; Waddington {et~al.} 2002; Bunker {et~al.} 2006).  Assuming passive evolution
of its stellar populations, \lbds\ lies well within the scatter of the
\markcite{kormendy:77}Kormendy (1977) relation, consistent with this milliJansky radio
source at $z \sim 1.5$ being an early-type galaxy which will passively
evolve into an ordinary $L^*$ elliptical by the present day.

\subsection{\lbdsb}

\lbdsb\ is a second, well-studied example of the evolved elliptical
subclass of EROs.  Slightly brighter than \lbds\ at optical and
near-infrared wavelengths, \lbdsb\ is a weak radio source whose
optical spectrum reveals an $\approx 4.0$~Gyr-old stellar population
at $z = 1.432$ \markcite{nolan:03, dey:06}(Nolan {et~al.} 2003; Dey {et~al.} 2006).  NICMOS images again reveal
a regular, $r^{1/4}$ profile, consistent with the Kormendy relation
\markcite{waddington:02, Bunker:06}(Waddington {et~al.} 2002; Bunker {et~al.} 2006).  Submillimeter observations of
\lbdsb\ obtained by \markcite{archibald:01}Archibald {et~al.} (2001) fail to detect this galaxy,
again consistent with \lbdsb\ no longer being actively involved in
star formation.

\section{Observations and Data Reduction\label{sec:observations}}
 
We obtained images from 3.6 $\mu$m to 160 $\mu$m of these prototypical
EROs with the {\it Spitzer Space Telescope} as part of guaranteed time
observations.  These observations employed both the Infrared Array
Camera \markcite{fazio:04a}(IRAC; Fazio {et~al.} 2004) and the Multiband Imaging Photometer
for {\it Spitzer} \markcite{rieke:04}(MIPS; Rieke {et~al.} 2004).

\subsection{IRAC}
 
Both \lbds\ and \lbdsb\ were observed with IRAC on 2004 February 17,
and \hr\ was observed on 2004 February 18.  Exposures with IRAC totaled
900~sec per target, obtained as nine 100~sec dithered frames using the
small-scale, cycling dither pattern.  Due to the higher background at
longer wavelengths, there are twice as many frames in IRAC channel 4
(8.0 $\mu$m), each with half the frame time.

We made minor modifications to the pipeline-processed data.  Beginning
with the basic calibrated data from the IRAC pipeline (version
S11.0.2), we corrected the ``pulldown'' and ``muxbleed'' associated with
well-exposed pixels in IRAC channels 1 (3.6 $\mu$m) and 2 (4.5 $\mu$m).
The empirically-derived muxbleed corrections were derived from a
large set of IRAC data taken with the same exposure time.  We then
reran the MOPEX\footnote{\tt http://ssc.spitzer.caltech.edu/postbcd/.}
mosaicing software with modified cosmic-ray rejection parameters and
an output pixel scale reduced by 50\%\ to improve the sampling of the
final images.  Source Extractor \markcite{bertin:96}(Bertin \& Arnouts 1996) was used to derive
photometry in $5\arcsec$ diameter apertures for most sources, which were
then corrected to total magnitudes assuming sources are unresolved at the
relatively coarse ($\sim 2\arcsec$) spatial resolution of {\it Spitzer}.
The exceptions were galaxies in close proximity to other sources,
e.g. \lbds.  For these galaxies, where source blending can compromise
accurate photometry, we measured fluxes in $2\arcsec$ diameter apertures,
which were then corrected to total magnitudes assuming the sources
were unresolved.

Fig.~1 presents the processed IRAC images.  For reference,
optical images obtained with the Wide-Field Planetary Camera~2
\markcite{trauger:94}(WFPC2; Trauger, Ballester, \& Burrows 1994) on
the {\it HST} are also presented.  These data, all obtained with the
F814W filter, are from the Multimission Archive at the Space Telescope
Science Institute\footnote{{\tt http://archive.stsci.edu/}}.

\subsection{MIPS}
 
Both \hr\ and \lbdsb\ were observed with MIPS on 2004 April 8, and
\lbds\ was observed on 2004 April 10.  The observations were performed
in MIPS small scale photometry mode with 30~sec $\times$ 5 cycles, 10~sec
$\times$ 10 cycles and 10~sec $\times$ 5 cycles at 24, 70 and 160~$\mu$m,
respectively. The 160~$\mu$m observations also incorporated a 1 $\times$
3 map, stepping by 75\%\ of the field of view. The total integration
times on source were $\sim$ 2100~sec at 24~$\mu$m, 1000~sec at 70~$\mu$m
and 300~sec at 160$\mu$m.

The 24~$\mu$m and 70~$\mu$m pipeline (version S11.4.0) basic calibrated
data (BCD) products were the starting point for the reductions. The
24~$\mu$m BCDs were then reprocessed to correct for flat fielding patterns
due to residue on the MIPS scan mirror, ``jailbar'' patterns caused by
bright cosmic rays artificially suppressing one or more of the four
readout channels, and gradients in the background subtraction due to
variation in either the dark current or the actual sky background. The
70~$\mu$m BCDs were column and time filtered to correct for cosmic
rays and response variations.  Details of 24~$\mu$m and 70~$\mu$m
artifacts can be found in the MIPS data handbook. The reprocessed
BCD frames were mosaiced together using MOPEX.  For 160~$\mu$m, the
default pipeline processed mosaic failed to detect any of the sources.
For this bandpass, we present $3 \sigma$ upper limits derived from
the {\it Spitzer} Sensitivity Performance Estimation Tool\footnote{\tt
http://ssc.spitzer.caltech.edu/tools/senspet/.} for medium background
level.  This is a conservative limit since all three sources were actually
observed at low background levels ($S_{\rm 24 \mu m} < 18$ MJy sr$^{-1}$).

Photometry at 24~$\mu$m and 70~$\mu$m was performed using a point source
fitting algorithm, whereby the positions of the source in the shorter
wavelength IRAC data was fed in as a prior. This has been shown to provide
accurate photometry as demonstrated in the GOODS datasets (Chary et al.,
in preparation).  The 70~$\mu$m photometry derived here was corrected
upwards by approximately 5\%\ to account for flux suppression by the
time and column filtering process.

The {\it Spitzer} photometry is presented in Tables~\ref{table.phot_hr10}
$-$ \ref{table.phot_53w069}, along with photometry at other wavelengths
derived from the literature.  Where no detection is made, $3 \sigma$
upper limits are given.  The broadband SEDs of all three EROs are
presented in Fig.~2.

\section{Spectral Energy Distributions\label{sec:sed}}

We next fit the broadband SEDs with galaxy template models with the
goal of deriving basic properties of the galaxies, such as stellar
mass and far-infrared luminosity.  The new data from {\it Spitzer}
have the capacity to significantly refine previous determinations
of these properties.  For the SED fitting, described below, we omit
some of the photometric points presented in Table~1.  First, where
{\it HST} imaging is available at similar wavelengths to ground-based
imaging (\eg $i$-band/F814W and $J$-band/F110W), we use only the higher
signal-to-noise ratio {\it HST} photometry.  Second, we do not include $<
3\sigma$ detections in our fits.  Finally, we omit the early (circa 1985)
CCD Palomar 200$''$ $gri$ imaging of the LBDS sources whose photometry
typically compare poorly with more recent, deeper photometry obtained
by {\it HST} and the Keck telescopes.

We fit \markcite{Bruzual:03}Bruzual \& Charlot (2003, BC03) model SEDs
to the optical thru mid-infrared ($\lambda_{\rm obs} \leq 8 \mu$m)
fluxes of these sources.  We adopt a Salpeter initial mass function
with solar metallicity templates, leaving the mass of the galaxy, dust
extinction, age of the starburst and $e$-folding time of the starburst
as free parameters.  Dust was assumed to be distributed in a screen.
Both starburst and galactic extinction laws were tried. The models were
redshifted and convolved through the relevant filter response/quantum
efficiency curves. A weighted, reduced, chi-square fit was performed to
the observed data points after multiplication with a model normalization
factor which represents the combined mass of stars and gas. The mass
fraction in stars was extracted from the corresponding BC03 model file
and multiplied by the normalization factor to derive the total mass in
stars.  Fig.~2 presents the best-fit BC03 models to all three EROs and
Table~\ref{table.bc03} presents the parameters of the best-fit models.
Fig.~\ref{fig.sed} presents the full SED, while Fig.~\ref{fig.sed_v2}
only shows the bluer wavelengths of the SED, where the emission is
dominated by starlight.

We find that the LBDS galaxies have stellar masses of $\sim 3 \times
10^{11} M_\odot$, while \hr\ has a stellar mass roughly twice as large.
Modulo differences in the assumed initial mass function, these values are
$4 - 8\, \times$ the average stellar mass of an $L_*$ galaxy in the local
universe \markcite{Cole:01}($M_* \sim 7 \times 10^{10} M_\odot$; Cole {et
al.} 2001) or of evolved $2 \simlt z \simlt 3$ infrared-selected galaxies
in the {\it Hubble} Deep Field - South \markcite{Labbe:05}($M_* \sim 8
\times 10^{10} M_\odot$; Labb\'e {et al.} 2005).  We find ages of a few
Gyr for the evolved galaxies, comparable to the more carefully determined
ages based on fitting the observed optical spectra \markcite{spinrad:97,
nolan:03}(Spinrad {et~al.} 1997; Nolan {et~al.} 2003).  Surprisingly,
our best-fit model to \hr\ has a comparable age, albeit with heavy
extinction, $A_V = 2.4$.

To fit the longer wavelength {\it Spitzer} and (sub)mm data for \hr,
we adopted the \markcite{Chary:01}Chary \& Elbaz (2001, CE01) range of far-infrared spectral
energy distributions. The CE01 templates provide a range of mid- and
far-infrared SEDs as a function of the far-infrared luminosity, $L_{\rm
FIR} = L (8 - 1000 \mu {\rm m})$.  They are empirically derived templates
based on observations of nearby infrared luminous galaxies and incorporate
components associated with continua from multi-temperature dust grains,
emission from polycyclic aromatic hydrocarbons, and absorption due
to silicate dust.  These templates have been shown to follow the
luminosity-luminosity correlations in {\it IRAS} passbands seen for
nearby galaxies. As with the BC03 fits to the stellar emission, the dust
emission has been redshifted and convolved through the relevant passbands.
We also attempted to fit the observed fluxes with a combination of
the CE01 templates and obscuration based on the \markcite{Li:02b}Li \& Draine (2002) model
which would most prominently introduce a spectral feature at 9.7$\mu$m
corresponding to silicate absorption. We find that a fit including the
\markcite{Li:02b}Li \& Draine (2002) model for dust obscuration provides a better fit to the
mid-infrared data points, although the derived $L_{\rm FIR} = 3 \times
10^{12} L_\odot$ remains essentially unchanged. The resultant obscuration
corresponds to $A_{V} \sim 11$~mag, which is significantly larger than
that seen in the optical/near-infrared. This suggests that HR10
has regions of heavily obscured star-formation which make it a
ULIRG.  With more accurate measurements of the Wien
side of the thermal dust emission, our derived far-infrared luminosity is
approximately 40\%\ that derived by \markcite{greve:03}Greve {et~al.} (2003) and \markcite{elbaz:02b}Elbaz {et~al.} (2002).
Fig.~\ref{fig.sed} presents the best-fit CE01 far-infrared template.
Since the CE01 templates are based on real galaxies and include stellar
populations, Fig.~\ref{fig.sed} shows the best-fit CE01 template only
at longer wavelengths ($\lambda_{\rm rest} \geq 5~\mu$m), stitched to
the best-fit BC03 stellar population template at shorter wavelengths.

It has been suggested that \hr\ is a scaled up version of Arp~220
\markcite{elbaz:02b}(Elbaz {et~al.} 2002).  We find that the mid-infrared and far-infrared
SED of \hr\ shows significant differences in comparison with Arp~220
(Fig.~\ref{fig.arp220}). The mid-infrared colors of \hr\ are better
fit by an SED including strong polycyclic aromatic hydrocarbon (PAH)
in emission and weak silicate in absorption. The derived extinction from
the silicate absorption is $A_V \sim 10$~mag. In contrast, Arp~220 has a
very strong silicate feature, indicative of $A_V > 50 - 80$ mags. Since
\hr\ is approximately twice as luminous than Arp~220 \markcite{elbaz:02b}(Elbaz {et~al.} 2002),
the strength of the silicate feature apparently does not scale with
far-infrared luminosity as would be naively expected. \hr\ also appears
to have a cooler color temperature for the far-infrared emission which
is more typical for the local ULIRG population than the SED exhibited
by Arp~220.

Since the photometry of \hr\ is well-fit with purely stellar populations
with dust, we infer that the energetics of this galaxy is dominated by
starlight.  The radio detection of \hr\ is fully consistent with the CE01
template, implying that it obeys the radio -- far-infrared correlation
observed in local star-forming galaxies \markcite{Yun:02}(Yun \& Carilli
2002).  If \hr\ does indeed host an AGN, the active nucleus apparently
plays a negligible role in the broadband emission from this galaxy.

Finally, to fit the radio portion of the LBDS SEDs, we fit the data
with a self-absorbed synchrotron emission spectrum.  Neglecting the
high-energy cut-off, this can be approximated by a parabola-shaped
curve in the log-log plane for frequencies lower than the frequency at
which the plasma optical depth becomes unity, $\nu_t$
\markcite{Andreani:02}(\eg Andreani {et~al.} 2002).  We adopt $\nu_t = 31.6$~MHz for both
sources and fit for the normalization and curvature of the radio
spectrum.  The results are plotted in Fig.~\ref{fig.sed} and illustrate
that the synchrotron emission which dominates the radio portion of the
LBDS SEDs has negligible contribution to the optical through
far-infrared emission for these sources.

\section{Separating Dusty Starbursts from Old Ellipticals
\label{sec:separating}}

It has been nearly a decade since the recognition that EROs constitute a
bimodal population of dusty, star-forming galaxies and quiescent, evolved
galaxies.  As EROs seem to be the likely $z \approx 1.5$ progenitors of
present day early-type galaxies \markcite{moustakas:02}(\eg Moustakas \& Somerville 2002), measuring
their relative fractions with confidence is an important measurement
for feedback into models of galaxy formation.

Deep spectroscopy can unambiguously separate the two ERO populations,
but is very expensive in telescope resources and typically suffers from
high incompleteness rates.  Independent spectroscopic samples of
$\approx 25$ EROs studied by \markcite{Cimatti:02b}Cimatti {et~al.} (2002b) and \markcite{Yan:04b}Yan, Thompson, \& Soifer (2004a)
both found roughly equal distributions of emission line galaxies and
purely absorption line galaxies.  These spectroscopic classes are
presumed associated with dusty starbursts and evolved early-type EROs,
respectively.  More recently, \markcite{Doherty:05}Doherty {et~al.} (2005) reported on a
spectroscopic sample of EROs nearly twice as large as the previous
studies, finding that 75\%\ of their sample have a dominant old stellar
population, but only 28\%\ of the sample is completely devoid of
evidence of recent star formation.  With typically one-third of the
observed EROs lacking sufficient signal to enter the spectroscopic
samples, the error bars on these population statistics are
substantial.

In principle, high resolution morphologies from {\it HST} would be useful
at distinguishing dusty starbursts from quiescent, old galaxies.  This
expectation is supported by Fig.~1, which shows that the dusty starburst
\hr\ has a disturbed morphology at observed $I$-band, while both \lbds\
and \lbdsb\ have regular, undisturbed morphologies.  From a deep {\it
HST} WFPC2 archival study of 115 EROs, \markcite{Yan:03}Yan \& Thompson
(2003) found that $30 \pm 5$\% of EROs are bulge-dominated, $64 \pm 7$\%
are disk-dominated, and only 6\% defy morphological classification.
However, from their spectroscopic study, \markcite{Yan:04b}Yan {et~al.}
(2004a) found very little correlation between spectroscopic classification
and morphological classification, with both emission line galaxies and
absorption line galaxies evenly comprised of bulge- and disk-dominated
galaxies.  Similarly, from a large, magnitude-limited sample of EROs in
the GOODS-South field, \markcite{Moustakas:04}Moustakas {et~al.} (2004)
found that the mean SEDs at $0.3 - 2.2 \mu$m are basically independent of
morphology.  The implication is that there is no robust way to distinguish
ERO subclasses based purely on imaging data shortward of $\approx 3 \mu$m.

However, {\it Spitzer} mid-infrared data enable the ERO subclasses to
be separated.  Fig.~\ref{fig.colz} presents the best-fit ERO SEDs (\S~4)
redshifted to $1 < z < 2$, the range at which most EROs are thought
to reside.  Despite previous claims that $J - K$ might be a useful
discriminant between the ERO subclasses \markcite{Pozzetti:00}(\eg Pozzetti \& Mannucci 2000),
Fig.~\ref{fig.colz} clearly demonstrates that ground-based near-infrared
data cannot robustly distinguish the subclasses.

% The $J - K$ colors of the ERO subclasses differ by only $\approx
% 0.5$~mag, which is significantly less than systematic errors between
% various methods of measuring photometry (e.g., compare the published
% F110W and $J$ band photometry of the LBDS sources in Table~1).

In theory, the best-fit SEDs suggest that IRAC data out to $8 \mu$m
might be useful at separating the populations.  While the quiescent,
evolved galaxies have SEDs which are falling longward of the $1.6 \mu$m
stellar peak, the star-forming SED rises into the mid- and far-infrared
due to the cold gas and dust associated with star formation.  Though the
models differ by $1 - 1.5$~mag in $K -$~[8.0] and [3.6]~$-$~[8.0] colors,
actual observations show more subtle color differences.  Not surprisingly,
the $100\times$ difference in the $24~\mu$m properties of the best-fit
SED provide the most robust separation of the star-forming and evolved
galaxy EROs.

\section{Conclusions\label{sec:conclusions}}

We present {\it Spitzer} IRAC and MIPS observations of three
well-studied EROs.  \hr\ is a dusty starburst class of EROs, while the
weak radio sources \lbds\ and \lbdsb\ are evolved, quiescent EROs.  The
IRAC data show the $1.6 \mu$m stellar peak in all three sources,
providing accurate derivations of basic properties of their stellar
populations.  Consistent with the bright near-infrared magnitudes, we
find all three galaxies are massive, with stellar masses $M_* = (3 - 6)
\times 10^{11} M_\odot$.  We also derive ages of a few Gyr for all
three galaxies, though \hr\ is found to be dustier and to have
proportionally more recent star formation.

Only \hr\ is detected by MIPS, illustrating that $24 \mu$m photometry
provides a robust discriminant between the two primary classes of EROs.
Blueward of $10 \mu$m, we find that both ERO classes are similar;
identifying the relative fractions of massive galaxies
at $z \sim 1.5$ which are old or dusty and starforming requires either
longer wavelength data or deep spectroscopy.  From a sample of EROs
in a 64 square arcmin area of the {\it Spitzer} First Look Survey,
\markcite{Yan:04c}Yan {et~al.} (2004b) find that approximately half of the ERO population is
detected to a $24 \mu$m flux limit of $40 \mu$Jy (3$\sigma$), suggesting
a roughly even distribution of starbursts and quiescent galaxies in the
ERO classification.  This breakdown is consistent with the courageous,
yet incomplete, spectroscopic studies \markcite{Cimatti:02, Yan:04b,
Doherty:05}(\eg Cimatti {et~al.} 2002a; Yan {et~al.} 2004a; Doherty {et~al.} 2005).  Since EROs are the likely progenitors of massive early-type
galaxies in the early universe, this suggests both that substantial
formation of early-type galaxies occurs at $z \simgt 2$, but that many
massive galaxies are also still actively forming stars at $z \simlt 2$.

% \acknowledgements 
% 
% This work is based on observations made with the {\it Spitzer Space
% Telescope}, which is operated by the Jet Propulsion Laboratory, California
% Institute of Technology, under a NASA contract.  Support was provided
% by NASA through an award issued by JPL/Caltech.

%\bibliographystyle{apj} 
%% \bibliography

%\bibliography{}

%% TABLE 1a
\begin{deluxetable}{lcccc}
%\tablenum{1a}
\tabletypesize{\small}
\tablecaption{Photometry of \hr}
\tablehead{
\colhead{Observed} &
\colhead{Rest} &
\colhead{} &
\colhead{Detector/} &
\colhead{} \\
\colhead{Wavelength} &
\colhead{Wavelength} &
\colhead{Flux Density} &
\colhead{Instrument} &
\colhead{Reference}}
\startdata
3620 \AA ($U$) & 1480 \AA       & $< 0.03~ \mu$Jy               & WHT & 1 \\
4400 \AA ($B$) & 1800 \AA       & $0.16 \pm 0.07~ \mu$Jy        & UH 2.2~m & 2,3 \\
4900 \AA ($G$) & 2010 \AA       & $0.09 \pm 0.04~ \mu$Jy        & WHT & 1 \\
5420 \AA ($V$) & 2220 \AA       & $< 0.14~ \mu$Jy               & WHT & 1 \\
6450 \AA ($R$) & 2640 \AA       & $0.21 \pm 0.08~ \mu$Jy        & WHT & 1 \\
8140 \AA (F814W)\tablenotemark{\dag} & 3340 \AA & $0.52 \pm 0.06~ \mu$Jy        & WFPC2/{\it HST} & 4 \\
8200 \AA ($I$) & 3360 \AA       & $0.58 \pm 0.22~ \mu$Jy        & WHT & 1 \\
8490 \AA ($I$) & 3480 \AA       & $0.41 \pm 0.13~ \mu$Jy        & UH 2.2~m & 2,3 \\
1.2 $\mu$m ($J$)\tablenotemark{\dag} & 4920 \AA & $6.4 \pm 2.1~ \mu$Jy          & UH 2.2~m & 2,3 \\
1.6 $\mu$m ($H$)\tablenotemark{\dag} & 6560 \AA & $14.8 \pm 3.6~ \mu$Jy         & UH 2.2~m & 2,3 \\
2.2 $\mu$m ($K$)\tablenotemark{\dag} & 9020 \AA & $27.7 \pm 0.6~ \mu$Jy         & Keck/NIRC & 2 \\
{\bf 3.6 {\boldmath $\mu$}m}\tablenotemark{\dag} & {\bf 1.5 {\boldmath $\mu$}m} & {\boldmath $65.06 \pm 6.53~ \mu$}{\bf Jy}           & {\bf IRAC}/{\boldmath $Spitzer$} & {\bf 5} \\
{\bf 4.5 {\boldmath $\mu$}m}\tablenotemark{\dag} & {\bf 1.8 {\boldmath $\mu$}m} & {\boldmath $80.76 \pm 8.11~ \mu$}{\bf Jy}           & {\bf IRAC}/{\boldmath $Spitzer$} & {\bf 5} \\
{\bf 5.8 {\boldmath $\mu$}m}\tablenotemark{\dag} & {\bf 2.4 {\boldmath $\mu$}m} & {\boldmath $59.24 \pm 6.41~ \mu$}{\bf Jy}           & {\bf IRAC}/{\boldmath $Spitzer$} & {\bf 5} \\
{\bf 8.0 {\boldmath $\mu$}m}\tablenotemark{\dag} & {\bf 3.3 {\boldmath $\mu$}m} & {\boldmath $57.43 \pm 5.96~ \mu$}{\bf Jy}           & {\bf IRAC}/{\boldmath $Spitzer$} & {\bf 5} \\
% 4.5 $\mu$m\tablenotemark{\dag} & 1.8 $\mu$m & $80.76 \pm 8.11~ \mu$Jy           & IRAC/{\it Spitzer} & 5 \\
% 5.8 $\mu$m\tablenotemark{\dag} & 2.4 $\mu$m & $59.24 \pm 6.41~ \mu$Jy           & IRAC/{\it Spitzer} & 5 \\
% 8.0 $\mu$m\tablenotemark{\dag} & 3.3 $\mu$m & $57.43 \pm 5.96~ \mu$Jy           & IRAC/{\it Spitzer} & 5 \\
12 $\mu$m\tablenotemark{\dag} & 4.9 $\mu$m  & $85 \pm 50~ \mu$Jy        & ISOCAM/{\it ISO}\tablenotemark{\ddag} & 6 \\
%12 $\mu$m & 4.9 $\mu$m  & $< 75$ mJy                   & {\it IRAS} & 4 \\
15 $\mu$m\tablenotemark{\dag} & 6.1 $\mu$m  & $203 \pm 62~ \mu$Jy       & ISOCAM/{\it ISO}\tablenotemark{\ddag} & 6 \\
{\bf 24 {\boldmath $\mu$}m}\tablenotemark{\dag} & {\bf 9.8 {\boldmath $\mu$}m} & {\boldmath $350 \pm 50~ \mu$}{\bf Jy}           & {\bf MIPS}/{\boldmath $Spitzer$} & {\bf 5} \\
%24 $\mu$m\tablenotemark{\dag} & 9.8 $\mu$m      & $350 \pm 50~ \mu$Jy   & MIPS/{\it Spitzer} & 5 \\
%25 $\mu$m & 10.2 $\mu$m & $< 60$ mJy                   & {\it IRAS} & 4 \\
%60 $\mu$m & 24.6 $\mu$m & $< 84$ mJy                   & {\it IRAS} & 4 \\
{\bf 70 {\boldmath $\mu$}m}\tablenotemark{\dag} & {\bf 28.7 {\boldmath $\mu$}m} & {\boldmath $5.6 \pm 2.0~$}{\bf mJy}           & {\bf MIPS}/{\boldmath $Spitzer$} & {\bf 5} \\
%70 $\mu$m\tablenotemark{\dag} & 28.7 $\mu$m     & $5.6 \pm 2.0$ mJy     & MIPS/{\it Spitzer} & 5 \\
90 $\mu$m & 36.9 $\mu$m & $< 120$ mJy                   & ISOPHOT/{\it ISO} & 6 \\
100 $\mu$m & 41 $\mu$m  & $< 270$ mJy\tablenotemark{\S}         & {\it IRAS} & 4 \\
{\bf 160 {\boldmath $\mu$}m}\tablenotemark{\dag} & {\bf 66 {\boldmath $\mu$}m} & {\boldmath $< 50~$}{\bf mJy}           & {\bf MIPS}/{\boldmath $Spitzer$} & {\bf 5} \\
%160 $\mu$m & 66 $\mu$m  & $<50$ mJy             & MIPS/{\it Spitzer} & 5 \\
170 $\mu$m & 70 $\mu$m & $< 120$ mJy                  & ISOPHOT/{\it ISO} & 6 \\
450 $\mu$m\tablenotemark{\dag} & 184 $\mu$m & $32.2 \pm 8.5$ mJy                & SCUBA/JCMT & 4 \\
850 $\mu$m & 348 $\mu$m & $8.7 \pm 1.6$ mJy             & SCUBA/JCMT & 7 \\
850 $\mu$m & 348 $\mu$m & $4.89 \pm 0.74$ mJy           & SCUBA/JCMT & 4 \\
850 $\mu$m\tablenotemark{\dag} & 348 $\mu$m & $8 \pm 2$ mJy                     & SCUBA/JCMT & 8 \\
1350 $\mu$m\tablenotemark{\dag} & 553 $\mu$m& $2.13 \pm 0.63$ mJy               & SCUBA/JCMT & 4 \\
3.6 cm\tablenotemark{\dag} & 1.5 cm             & $35 \pm 11~ \mu$Jy            & \nodata & 9 \\
21 cm & 8.2 cm          & $< 300~ \mu$Jy\tablenotemark{\S}      & \nodata & 4 \\
\enddata

\tablenotetext{\dag}{Data used in SED modeling (\S 4).}
\tablenotetext{\ddag}{Systematic and statistical errors have been added in quadrature.}
\tablenotetext{\S}{Reference does not specify significance of non-detection.}
\tablerefs{(1) \markcite{haynes:02}Haynes {et~al.} (2002), (2) \markcite{graham:96}Graham \& Dey (1996), (3)
\markcite{hu:94}Hu \& Ridgway (1994), (4) \markcite{dey:99}Dey {et~al.} (1999), (5) This paper, (6) \markcite{elbaz:02b}Elbaz {et~al.} (2002),
(7) \markcite{cimatti:98b}Cimatti {et~al.} (1998), (8) \markcite{greve:03}Greve {et~al.} (2003), (9) \markcite{frayer:96}Frayer (1996)}

\label{table.phot_hr10}
\end{deluxetable}

%% TABLE 1b
\begin{deluxetable}{lcccc}
%\tablenum{1b}
\tablecaption{Photometry of \lbds}
\tablehead{
\colhead{Observed} &
\colhead{Rest} &
\colhead{} &
\colhead{Detector/} &
\colhead{} \\
\colhead{Wavelength} &
\colhead{Wavelength} &
\colhead{Flux Density} &
\colhead{Instrument} &
\colhead{Reference}}
\startdata
4960 \AA ($g$) & 1940 \AA       & $0.16 \pm 0.07~ \mu$Jy        & Palomar 200$''$ & 1 \\
6420 \AA ($R$)\tablenotemark{\dag} & 2520 \AA   & $0.47 \pm 0.09~ \mu$Jy        & Keck/LRIS & 2 \\
6480 \AA ($r$) & 2540 \AA       & $0.14 \pm 0.06~ \mu$Jy        & Palomar 200$''$ & 1 \\
8140 \AA (F814W)\tablenotemark{\dag} & 3190 \AA & $0.80 \pm 0.02~ \mu$Jy        & WFPC2/{\it HST} & 3 \\
8160 \AA ($i$) & 3200 \AA       & $0.45 \pm 0.08~ \mu$Jy        & Palomar 200$''$ & 1 \\
1.1 $\mu$m (F110W)\tablenotemark{\dag} & 4310 \AA       & $3.44 \pm 0.10~ \mu$Jy        & NICMOS/{\it HST} & 3 \\ 
1.2 $\mu$m ($J$) & 4700 \AA     & $10.00 \pm 0.92~ \mu$Jy       & UKIRT & 1 \\
1.6 $\mu$m ($H$)\tablenotemark{\dag} & 6270 \AA & $15.85 \pm 1.46~ \mu$Jy       & UKIRT & 1 \\
2.2 $\mu$m ($K$)\tablenotemark{\dag} & 8620 \AA & $21.48 \pm 2.58~ \mu$Jy       & UKIRT & 1 \\
{\bf 3.6 {\boldmath $\mu$}m}\tablenotemark{\dag} & {\bf 1.4 {\boldmath $\mu$}m} & {\boldmath $35.63 \pm 3.64~ \mu$}{\bf Jy}           & {\bf IRAC}/{\boldmath $Spitzer$} & {\bf 4} \\
{\bf 4.5 {\boldmath $\mu$}m}\tablenotemark{\dag} & {\bf 1.8 {\boldmath $\mu$}m} & {\boldmath $37.22 \pm 3.82~ \mu$}{\bf Jy}           & {\bf IRAC}/{\boldmath $Spitzer$} & {\bf 4} \\
{\bf 5.8 {\boldmath $\mu$}m}\tablenotemark{\dag} & {\bf 2.3 {\boldmath $\mu$}m} & {\boldmath $19.98 \pm 3.01~ \mu$}{\bf Jy}           & {\bf IRAC}/{\boldmath $Spitzer$} & {\bf 4} \\
{\bf 8.0 {\boldmath $\mu$}m}\tablenotemark{\dag} & {\bf 3.1 {\boldmath $\mu$}m} & {\boldmath $14.23 \pm 2.04~ \mu$}{\bf Jy}           & {\bf IRAC}/{\boldmath $Spitzer$} & {\bf 4} \\
% 3.6 $\mu$m\tablenotemark{\dag} & 1.4 $\mu$m & $35.63 \pm 3.64~ \mu$Jy   & IRAC/{\it Spitzer} & 4 \\
% 4.5 $\mu$m\tablenotemark{\dag} & 1.8 $\mu$m & $37.22 \pm 3.82~ \mu$Jy   & IRAC/{\it Spitzer} & 4 \\
% 5.8 $\mu$m\tablenotemark{\dag} & 2.3 $\mu$m & $19.98 \pm 3.01~ \mu$Jy   & IRAC/{\it Spitzer} & 4 \\
% 8.0 $\mu$m\tablenotemark{\dag} & 3.1 $\mu$m & $14.23 \pm 2.04~ \mu$Jy   & IRAC/{\it Spitzer} & 4 \\
{\bf 24 {\boldmath $\mu$}m}\tablenotemark{\dag} & {\bf 9.4 {\boldmath $\mu$}m} & {\boldmath $< 40~ \mu$}{\bf Jy}           & {\bf MIPS}/{\boldmath $Spitzer$} & {\bf 4} \\
% 24 $\mu$m & 9.4 $\mu$m  & $<40~ \mu$Jy                  & MIPS/{\it Spitzer} & 4 \\
{\bf 70 {\boldmath $\mu$}m}\tablenotemark{\dag} & {\bf 27 {\boldmath $\mu$}m} & {\boldmath $< 8.1~$}{\bf mJy}           & {\bf MIPS}/{\boldmath $Spitzer$} & {\bf 4} \\
% 70 $\mu$m & 27 $\mu$m  & $<8.1$ mJy                     & MIPS/{\it Spitzer} & 4 \\
{\bf 160 {\boldmath $\mu$}m}\tablenotemark{\dag} & {\bf 63 {\boldmath $\mu$}m} & {\boldmath $< 50~$}{\bf mJy}           & {\bf MIPS}/{\boldmath $Spitzer$} & {\bf 4} \\
% 160 $\mu$m & 63 $\mu$m  & $<50$ mJy                     & MIPS/{\it Spitzer} & 4 \\
6.2 cm\tablenotemark{\dag} & 2.4 cm             & $6.5 \pm 0.4$ mJy             & \nodata & 2 \\
19 cm\tablenotemark{\dag} & 7.4 cm              & $23.0 \pm 1.7$ mJy            & \nodata & 2 \\
21 cm\tablenotemark{\dag} & 8.2 cm              & $22.1 \pm 2.0$ mJy            & \nodata & 1 \\
50 cm\tablenotemark{\dag} & 20 cm               & $66.0 \pm 3.9$ mJy            & \nodata & 1 \\
\enddata

\tablenotetext{\dag}{Data used in SED modeling (\S 4).}
\tablerefs{(1) \markcite{waddington:00}Waddington {et~al.} (2000), (2) \markcite{spinrad:97}Spinrad {et~al.} (1997), (3)
\markcite{waddington:02}Waddington {et~al.} (2002), (4) This paper}

\label{table.phot_53w091}
\end{deluxetable}

%% TABLE 1c
\begin{deluxetable}{lcccc}
%\tablenum{1c}
\tablecaption{Photometry of \lbdsb}
\tablehead{
\colhead{Observed} &
\colhead{Rest} &
\colhead{} &
\colhead{Detector/} &
\colhead{} \\
\colhead{Wavelength} &
\colhead{Wavelength} &
\colhead{Flux Density} &
\colhead{Instrument} &
\colhead{Reference}}
\startdata
4960 \AA ($g$) & 2040 \AA       & $0.10 \pm 0.06~ \mu$Jy        & Palomar 200$''$ & 1 \\
6480 \AA ($r$) & 2660 \AA       & $0.30 \pm 0.06~ \mu$Jy        & Palomar 200$''$ & 1 \\
8140 \AA (F814W)\tablenotemark{\dag} & 3350 \AA & $0.77 \pm 0.02~ \mu$Jy        & WFPC2/{\it HST} & 2 \\
8160 \AA ($i$) & 3360 \AA       & $0.46 \pm 0.06~ \mu$Jy        & Palomar 200$''$ & 1 \\
1.1 $\mu$m (F110W)\tablenotemark{\dag} & 4520 \AA       & $3.80 \pm 0.07~ \mu$Jy        & NICMOS/{\it HST} & 2 \\ 
1.2 $\mu$m ($J$) & 4930 \AA     & $12.59 \pm 1.63~ \mu$Jy       & UKIRT & 1 \\
1.6 $\mu$m ($H$)\tablenotemark{\dag} & 6580 \AA & $12.59 \pm 2.10~ \mu$Jy       & UKIRT & 1 \\
2.2 $\mu$m ($K$)\tablenotemark{\dag} & 9050 \AA & $24.43 \pm 2.48~ \mu$Jy       & UKIRT & 1 \\
{\bf 3.6 {\boldmath $\mu$}m}\tablenotemark{\dag} & {\bf 1.5 {\boldmath $\mu$}m} & {\boldmath $43.48 \pm 4.38~ \mu$}{\bf Jy}           & {\bf IRAC}/{\boldmath $Spitzer$} & {\bf 3} \\
{\bf 4.5 {\boldmath $\mu$}m}\tablenotemark{\dag} & {\bf 1.9 {\boldmath $\mu$}m} & {\boldmath $43.13 \pm 4.35~ \mu$}{\bf Jy}           & {\bf IRAC}/{\boldmath $Spitzer$} & {\bf 3} \\
{\bf 5.8 {\boldmath $\mu$}m}\tablenotemark{\dag} & {\bf 2.4 {\boldmath $\mu$}m} & {\boldmath $20.72 \pm 2.92~ \mu$}{\bf Jy}           & {\bf IRAC}/{\boldmath $Spitzer$} & {\bf 3} \\
{\bf 8.0 {\boldmath $\mu$}m}\tablenotemark{\dag} & {\bf 3.3 {\boldmath $\mu$}m} & {\boldmath $22.18 \pm 2.70~ \mu$}{\bf Jy}           & {\bf IRAC}/{\boldmath $Spitzer$} & {\bf 3} \\
% 3.6 $\mu$m\tablenotemark{\dag} & 1.5 $\mu$m & $43.48 \pm 4.38~ \mu$Jy   & IRAC/{\it Spitzer} & 3 \\
% 4.5 $\mu$m\tablenotemark{\dag} & 1.9 $\mu$m & $43.13 \pm 4.35~ \mu$Jy   & IRAC/{\it Spitzer} & 3 \\
% 5.8 $\mu$m\tablenotemark{\dag} & 2.4 $\mu$m & $20.72 \pm 2.92~ \mu$Jy   & IRAC/{\it Spitzer} & 3 \\
% 8.0 $\mu$m\tablenotemark{\dag} & 3.3 $\mu$m & $22.18 \pm 2.70~ \mu$Jy   & IRAC/{\it Spitzer} & 3 \\
{\bf 24 {\boldmath $\mu$}m}\tablenotemark{\dag} & {\bf 9.9 {\boldmath $\mu$}m} & {\boldmath $< 50~ \mu$}{\bf Jy}           & {\bf MIPS}/{\boldmath $Spitzer$} & {\bf 3} \\
% 24 $\mu$m & 9.9 $\mu$m  & $<50~ \mu$Jy                  & MIPS/{\it Spitzer} & 3 \\
{\bf 70 {\boldmath $\mu$}m}\tablenotemark{\dag} & {\bf 29 {\boldmath $\mu$}m} & {\boldmath $< 7.8~$}{\bf mJy}           & {\bf MIPS}/{\boldmath $Spitzer$} & {\bf 3} \\
% 70 $\mu$m & 29 $\mu$m  & $<7.8$ mJy                     & MIPS/{\it Spitzer} & 3 \\
{\bf 160 {\boldmath $\mu$}m}\tablenotemark{\dag} & {\bf 66 {\boldmath $\mu$}m} & {\boldmath $< 50~$}{\bf mJy}           & {\bf MIPS}/{\boldmath $Spitzer$} & {\bf 3} \\
% 160 $\mu$m & 66 $\mu$m  & $<50$ mJy                     & MIPS/{\it Spitzer} & 3 \\
450 $\mu$m & 185 $\mu$m & $<49$ mJy                     & SCUBA/JCMT & 4 \\
850 $\mu$m & 360 $\mu$m & $<3.12$ mJy                   & SCUBA/JCMT & 4 \\
21 cm\tablenotemark{\dag} & 8.6 cm              & $3.7 \pm 0.3$ mJy             & \nodata & 1 \\
50 cm\tablenotemark{\dag} & 21 cm               & $7.8 \pm 0.9$ mJy             & \nodata & 1 \\
\enddata

\tablenotetext{\dag}{Data used in SED modeling (\S 4).}
\tablerefs{(1) \markcite{waddington:00}Waddington {et~al.} (2000), (2) \markcite{waddington:02}Waddington {et~al.} (2002), (3)
This paper, (4) \markcite{archibald:01}Archibald {et~al.} (2001)}

\label{table.phot_53w069}
\end{deluxetable}

%% TABLE 2
\begin{deluxetable}{lccc}
%\tablenum{2}
\tablecaption{BC03 Model Fits to ERO Stellar Populations}
\tablehead{
\colhead{} &
\colhead{\hr} &
\colhead{\lbds} &
\colhead{\lbdsb}}
\startdata
$M_*$ [$10^{11} M_\odot$] & 5.6 & 3.0 & 3.2 \\
$A_V$ [mag]               & 2.4 & 0.4 & 1.4 \\
age [Gyr]                 & 3.0 & 3.5 & 4.0 \\
$e$-folding [Gyr]         & 2.0 & 0.7 & 2.0 \\
$L_{\rm FIR}$ [$L_\odot$] & $3\times10^{12}$ & \nodata & \nodata \\
\enddata

\label{table.bc03}
\end{deluxetable}

% FIGURE 1a

% FIGURE 1b
\begin{figure}
\figurenum{1b}
\begin{center}
\plotfiddle{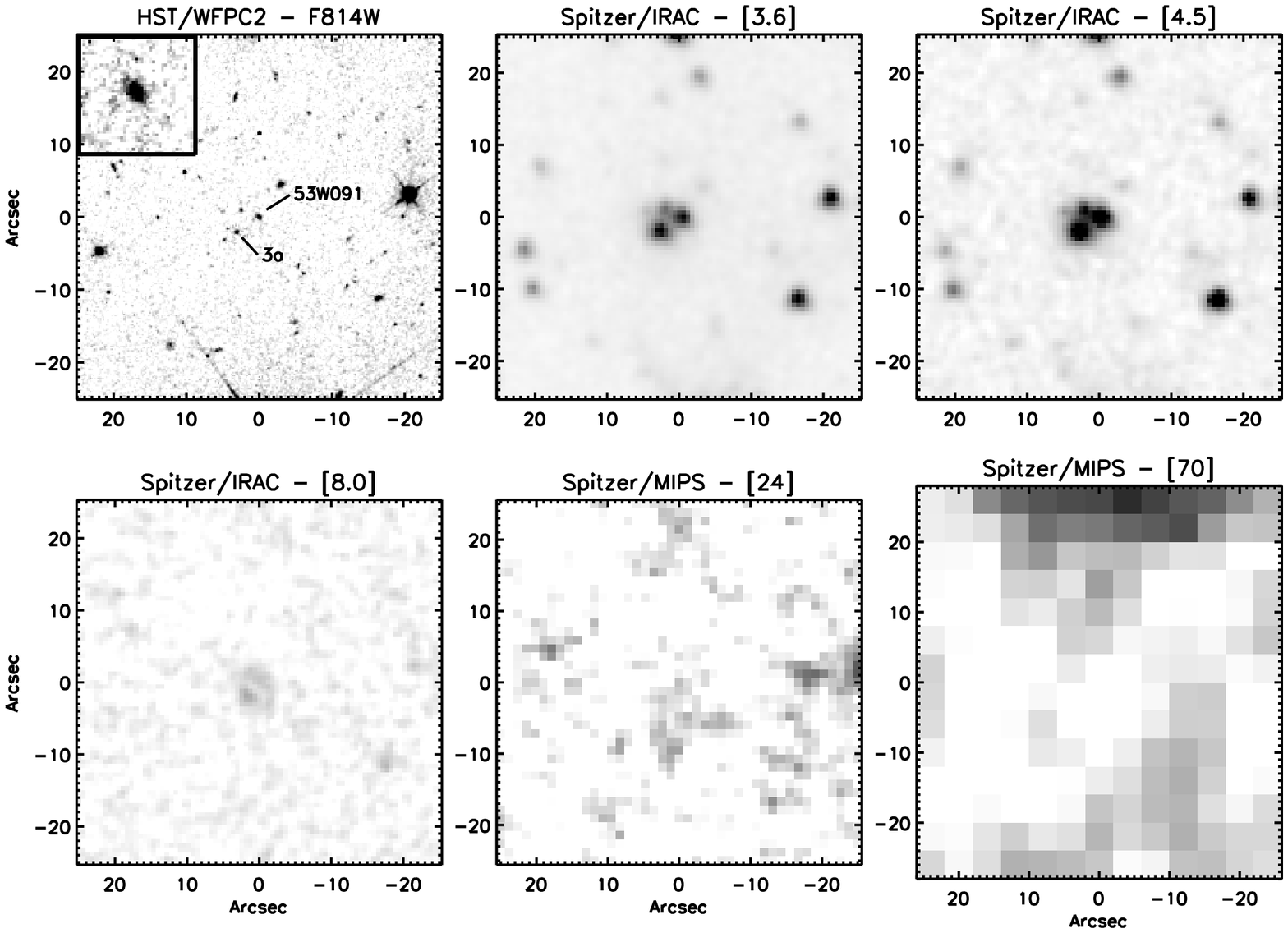}{4.2in}{0}{75}{75}{-240}{0}
\end{center}

\caption{{\it HST} and {\it Spitzer} images of the prototypical ERO
\lbds, located at $\alpha = 17^{\rm h}22^{\rm m}32.63^{\rm s}, \delta =
+50\deg06\arcmin01\farcs5$ (J2000).  Cameras and filters are indicated.
The fields of view shown are 50\arcsec\ on a side and relevant sources are
labeled.  North is up, and east is to the left.  The inset is 4\arcsec\
on side.  \lbds\ is not detected in the MIPS bands.}

\label{fig.image_53w091}
\end{figure}

% FIGURE 1c
\begin{figure}
\figurenum{1c}
\begin{center}
\plotfiddle{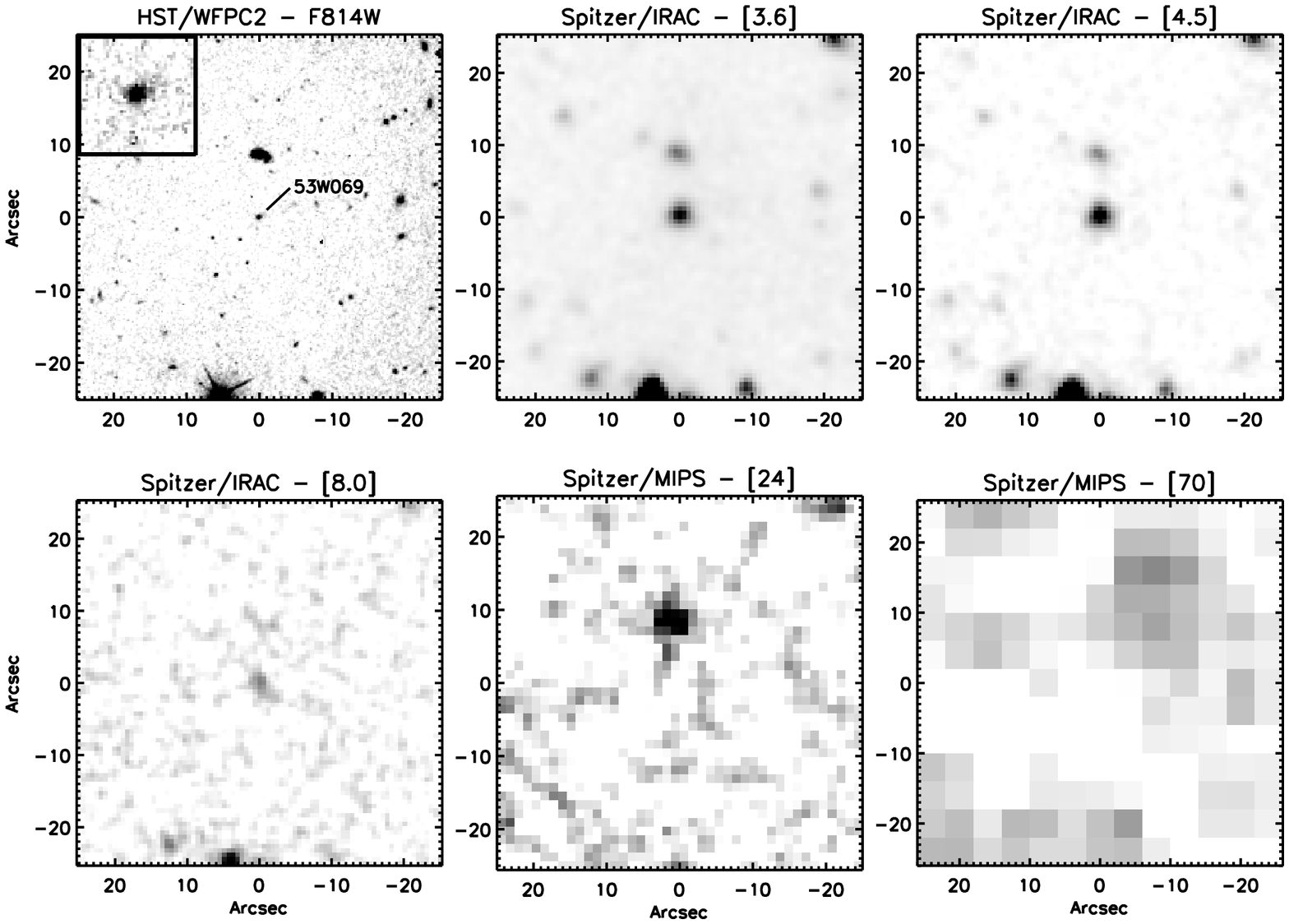}{4.2in}{0}{75}{75}{-240}{0}
\end{center}

\caption{{\it HST} and {\it Spitzer} images of the prototypical ERO
\lbdsb, located at $\alpha = 17^{\rm h}20^{\rm m}02.5^{\rm s}, \delta =
+49\deg44\arcmin51\arcsec$ (J2000).  Cameras and filters are indicated.
The fields of view shown are 50\arcsec\ on a side.  North is up, and
east is to the left.  The inset is 4\arcsec\ on side.  \lbdsb\ is not
detected in the MIPS bands.}

\label{fig.image_53w069}
\end{figure}

% FIGURE 2a
\begin{figure}[!t]
\figurenum{2a}
\begin{center}
\plotfiddle{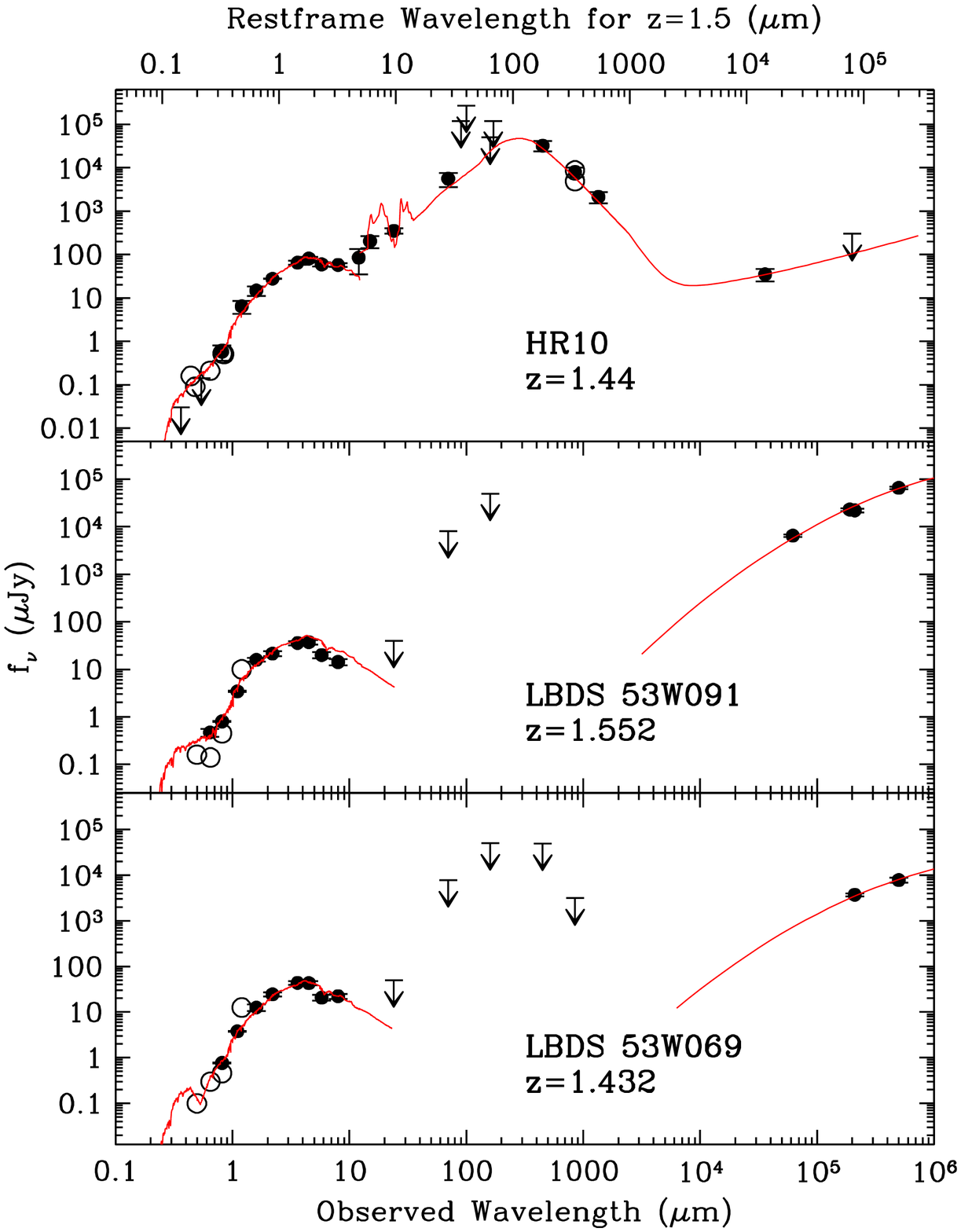}{7.6in}{0}{80}{80}{-250}{-40}
\end{center}
                                                                                          
\caption{SEDs of \hr, \lbds, and \lbdsb.  The topmost axis indicates
restframe wavelengths for $z = 1.5$, the approximate redshift of all
three galaxies.  Photometry not used in the SED fits are indicated with
open symbols (see \S 4).  The solid line shows the best-fit SEDs.}

\label{fig.sed}
\end{figure}

% FIGURE 2b
\begin{figure}[!t]
\figurenum{2b}
\begin{center}
\plotfiddle{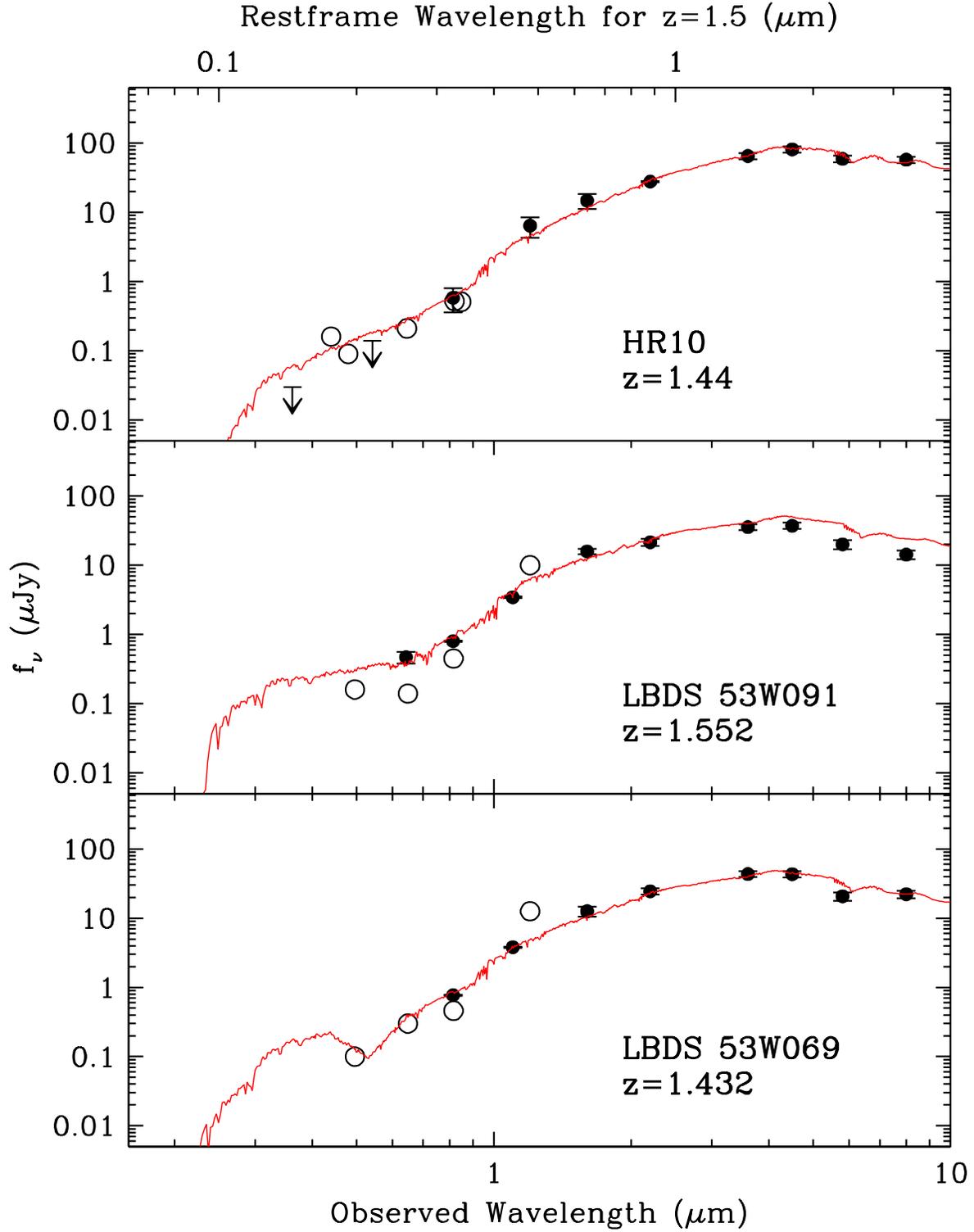}{7.6in}{0}{80}{80}{-250}{-40}
\end{center}
                                                                                          
\caption{SEDs of \hr, \lbds, and \lbdsb, restricted to the stellar portion
of the spectrum.  The topmost axis indicates restframe wavelengths for
$z = 1.5$, the approximate redshift of all three galaxies.  Photometry
not used in the SED fits are indicated with open symbols (see \S 4).
The solid line shows the best-fit SEDs.}

\label{fig.sed_v2}
\end{figure}

% FIGURE 3
\begin{figure}[!t]
\figurenum{3}
\begin{center}
\plotfiddle{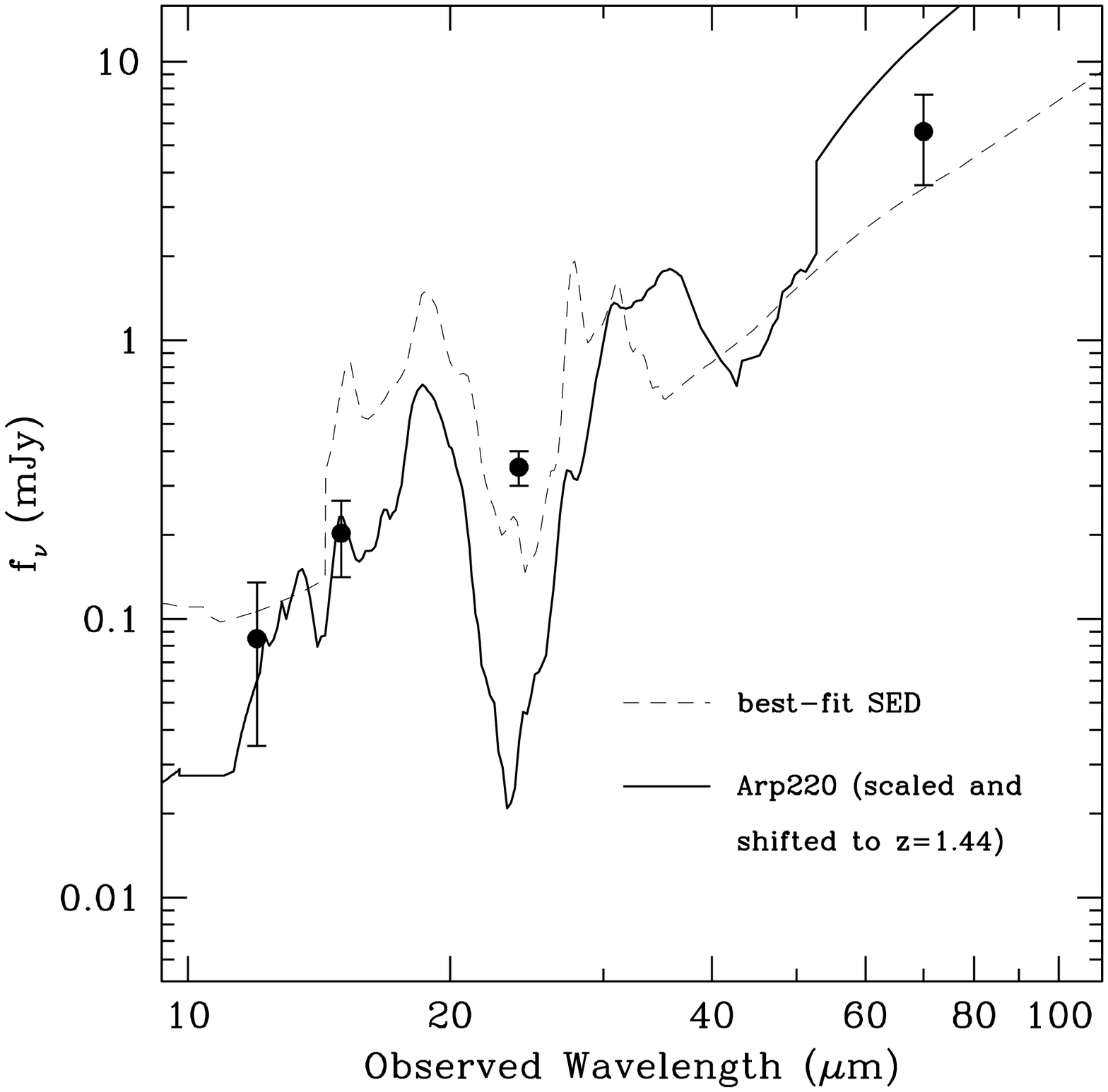}{2.5in}{0}{60}{60}{-180}{-80}
\end{center}

\caption{Mid-infrared spectrum of Arp~220, shifted to $z = 1.44$ and
scaled to approximate the $\lambda < 20 \mu$m photometry for \hr\ (filled
circles).  \hr\ has significantly less silicate absorption than Arp~220
and appears to have cooler dust dominating the far-infrared emission.}

\label{fig.arp220}
\end{figure}

% FIGURE 4
\begin{figure}[!t]
\figurenum{4}
\begin{center}
\plotfiddle{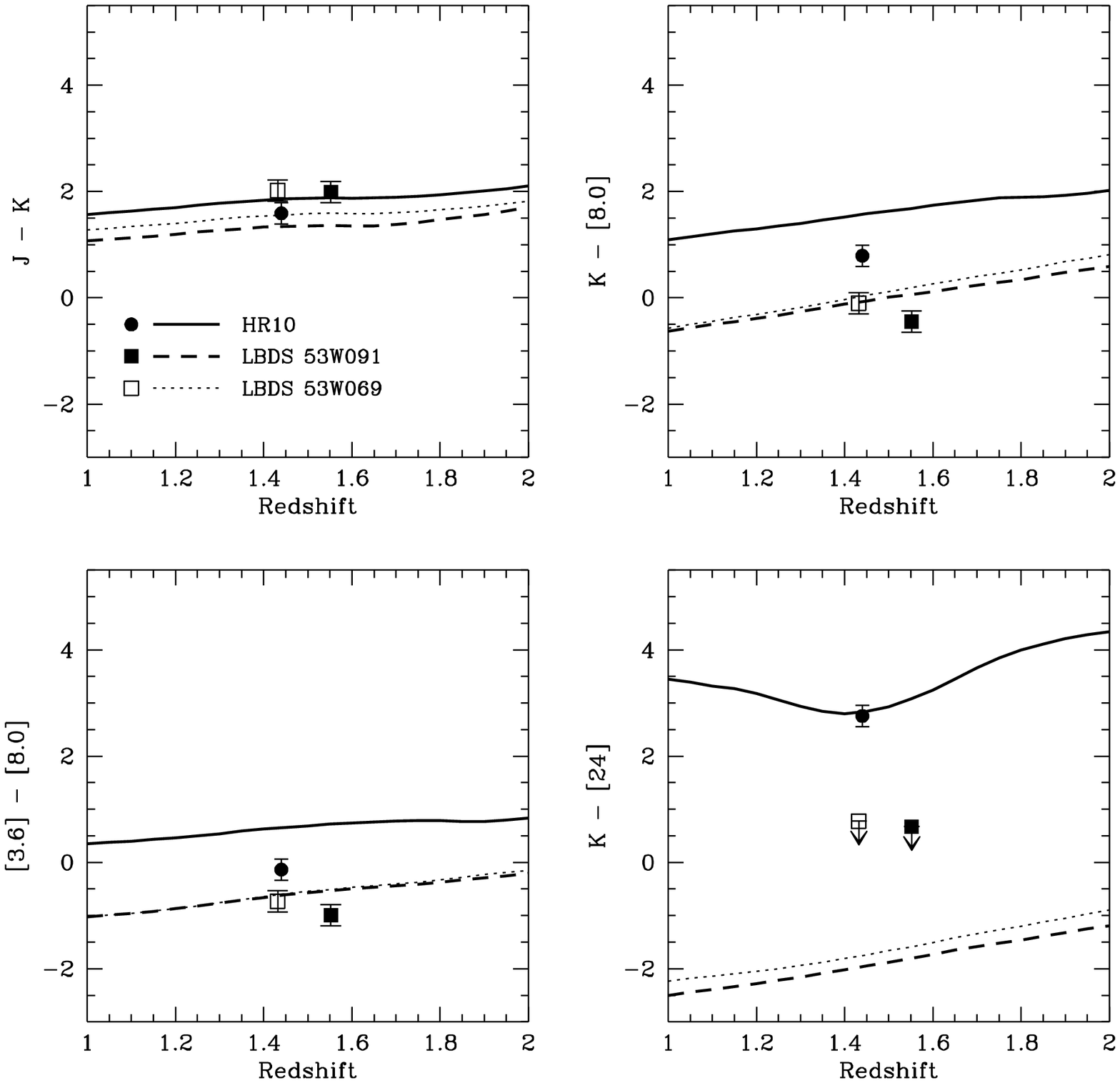}{4.2in}{0}{80}{80}{-240}{-140}
\end{center}
                                                                                              
\caption{Infrared colors (in the AB system) for the best-fit SEDs
redshifted to $1 < z < 2$, the expected redshift range of EROs.  A
Rayleigh-Jeans tail has been assumed for the LBDS galaxies at
$\lambda_{\rm obs} > 25 \mu$m.  Colors of the prototypical EROs are
also shown with uniform 20\%\ error bars.  Though the models suggest
that IRAC $8 \mu$m data might be effective at separating dusty
starbursts (e.g., \hr) from evolved stellar populations (e.g., \lbds\,
and \lbdsb), actual photometry implies that MIPS $24 \mu$m data is
necessary.}

\label{fig.colz} 
\end{figure}

\clearpage
\end{document}